\documentclass[12pt,showpacs,preprintnumbers,superscriptaddress,amsmath,amssymb,nofootinbib]{revtex4-1}
\usepackage{graphicx,hyperref}% Include figure files
\usepackage{dcolumn}% Align table columns on decimal point
\usepackage{bm}% bold math
\usepackage{amssymb}
\usepackage{amsmath}
\usepackage{epsfig}    
\usepackage{color}
\usepackage{slashed}
\usepackage{hhline}
\def\be{\begin{equation}}
\def\ee{\end{equation}}
\newcommand{\bea}{\begin{eqnarray}}
\newcommand{\eea}{\end{eqnarray}}
\newcommand{\Z}{\mathbb{Z}}

\numberwithin{equation}{section}

\begin{document}

%%%%%%%%%
\title{ 3.5 keV X-ray Line Signal from Decay of Right-Handed Neutrino due to Transition Magnetic Moment
}

\author{Kamakshya Prasad Modak}
\email{kamakshya.modak@saha.ac.in}
\affiliation{Astroparticle Physics and Cosmology Division, Saha Institute of Nuclear Physics, Kolkata 700064, India}

%\vspace{2.0cm}

\begin{abstract}

We consider the dark matter model with radiative neutrino mass generation
where the Standard Model is extended with three right-handed singlet neutrinos 
($N_1$, $N_2$ and $N_3$) and one additional SU(2)$_L$ doublet scalar $\eta$.
One of the right-handed neutrinos ($N_1$), being lightest among them, is
a leptophilic fermionic dark matter candidate whose stability
is ensured by the imposed $\Z_2$ symmetry on this model. 
%The neutrino masses are generated radiatively.
The second lightest right-handed neutrino ($N_2$) is assumed to be nearly
degenerated with the lightest one enhancing the co-annihilation
between them.
The effective interaction term among the lightest, second lightest
right-handed neutrinos and photon containing transition magnetic moment
is responsible for the decay of heavier right-handed neutrino 
to the lightest one and a photon ($N_2\to N_1 + \gamma$).
This radiative decay of heavier right-handed neutrino %to the the lightest one
with charged scalar and leptons in internal lines
could explain the X-ray line signal $\sim$ $3.5$ keV
recently claimed by XMM-Newton X-ray observatory
from different galaxy clusters and Andromeda galaxy (M31).
The value of the transition magnetic moment is computed and found
to be several orders of magnitude below the current reach of various direct
dark matter searches.
The other parameter space in this framework in the light of the 
observed signal is further investigated.

\end{abstract}
\maketitle
\newpage

\section{Introduction}
%%%%%%%%%%%%%%%%%%%%%%%%%%%%%%%%%%%%%%

One of the enigmas of modern particle physics is dark matter (DM) which, according to the recent
survey of PLANCK~\cite{Ade:2013lta}, consists of $\sim$ 26.8\% of the total energy content of the universe.
Various astrophysical and cosmological observations~\cite{Begeman:1991iy, Komatsu:2010fb, Massey:2007wb} strongly suggest 
convincing hints of the
existence of dark matter which is non-relativistic or cold in nature. 
The particle nature of dark matter is still unknown.
The weakly interacting massive particles (WIMPs) are the most promising candidates
for cold dark matter. 

The experimental techniques for the detection of dark matter for both direct and indirect cases
are very challenging. In direct detection experiments, the recoil energy of the target nucleus
scattered off by DM particle is measured whereas the signatures of the annihilations of decays
of DM particles such as charged particles, photons and neutrinos etc. are aimed to detect in indirect
searches. The monochromatic line feature of such decay or annihilation products of DM are
particularly significant in predicting the nature of DM particles.
A huge variety of DM models in the framework of WIMP scenario
with masses of DM spanning from keV to TeVs has been addressed in several 
literatures and their
direct and indirect detection prospects have been widely 
studied~\cite{scalar_singlet, veltman, Burgess:2000yq, kaluza, triplet, Modak:2012wk, smssm, axion, 
sing_ferm, idm, Cirelli, Hooper:2012sr, Modak:2013jya}.

Recently an evidence of X-ray line of energy 3.55 keV with more than $3\sigma$ CL has been reported from the
analysis of X-ray data of 73 galaxy clusters from XMM-Newton observatory~\cite{Bulbul:2014sua}. Another group
has also claimed a similar line (3.52 keV X-ray line at $4.4\sigma$ CL) from the data of X-ray spectra
of Andromeda galaxy (M31) and Perseus cluster~\cite{Boyarsky:2014jta}. The galaxy clusters 
are assumed to contain huge amount of DM. Thus the signal may have a possible origin
related to DM. The observed line has been explained as decay of sterile
neutrino dark matter ($\nu_s\to \nu + \gamma$) with mass of the sterile neutrino $7.06\pm0.05$ keV
and mixing angle $\sin^2(2\theta) = (2.2-20)\times 10^{-11}$~\cite{Boyarsky:2014jta}. 
Recently many other interesting ideas have been proposed to explain this line
signal to come from DM~\cite{Ishida:2014dlp, Finkbeiner:2014sja, 
Higaki:2014zua, Jaeckel:2014qea,Lee:2014xua, Kong:2014gea, Choi:2014tva, Baek:2014qwa, Tsuyuki:2014aia,
Bezrukov:2014nza, Kolda:2014ppa, Allahverdi:2014dqa, Queiroz:2014yna, Babu:2014pxa, Dudas:2014ixa,
Demidov:2014hka, Ko:2014xda, Bomark:2014yja, Liew:2014gia, Krall:2014dba, Aisati:2014nda, Frandsen:2014lfa}. 

The neutrino oscillation data~\cite{Fukuda:1998mi, Ahmad:2002jz, Araki:2004mb, Adamson:2008zt}
provide strong evidences for neutrino mass. 
The non-zero neutrino masses and evidences of DM give hints to the physics beyond the Standard Model (SM).
The two beyond SM phenomenon, namely the origin of neutrino masses and 
the existence of cold dark matter may have a connection.
In this work we focus on the simplest framework which invokes this idea of connecting both sectors has been proposed by Ma~\cite{ma}.
In this model the neutrino masses are generated via radiative processes with only the DM particles
in the loop. The right-handed neutrino which can be a possible DM candidate interacts with lepton doublets 
and hence DM in this scenario is leptophilic in nature. The imposed 
discreet $\Z_2$ symmetry on this model not only forbids the tree-level Dirac mass terms but also 
assure a stable cold DM candidate. Phenomenological prospects for DM in this model have been done in
Refs.~\cite{Kubo:2006yx,
  Kajiyama:2006ww, Suematsu:2009ww, Suematsu:2010gv, Sierra:2008wj,
  Kajiyama:2011fe, Kajiyama:2011fx}. 
In this paper we consider the case where the lightest right-handed neutrino ($N_1$) is the cold DM candidate and
the second lightest right-handed neutrino ($N_2$) is nearly degenerated with the cold DM candidate. This
situation provides rich phenomenology in direct detection of such dark matter candidate~\cite{Schmidt:2012yg}. Elastic 
scattering cross section for DM-nucleon interaction is suppressed in this case and inelastic
scattering that occurs radiatively dominates.
The transition from $N_2$ to $N_1$ gives rise to monochromatic photon
with energy equal to the mass difference between the lightest and second lightest 
right-handed neutrinos. If the mass difference between $N_2$ and $N1$ is of $\sim$ keV,
then the recent observation of X-ray line can be accommodated in this beyond SM scenario. 

The paper is organised as follows. In Sec. I the theoretical framework of the model 
is briefly discussed. Explanation of the observed X-ray line in this model framework
and a study of the constrained parameter space are done in the next section. 
In Sec. IV a brief summery of this work and some conclusions
are drawn.

%%%%%%%%%%%%%%%%%%%%%%%%%%%%%%%%%%%%%%
\section{The Model}

We consider the model proposed by Ma~\cite{ma} which is the extension of Standard Model with three
gauge singlet right-handed neutrinos $N_1$, $N_2$, $N_3$ and and extra SU(2)$_L$ doublet
scalar $\eta$. 
The fields can be written as,
 \begin{eqnarray}
N_1,\quad N_2,\quad N_3,\quad \eta &=\left(
 \begin{array}{c}
  \eta^{+} \\
 \eta^{0}\\
 \end{array}
 \right).
 \end{eqnarray}
The doublet scalar $\eta$ is assumed to obtain no vacuum expectation value
and hence inert. An additional discreet $\Z_2$ symmetry is imposed on the model. The stability
of the cold dark matter candidate in this model is guaranteed by this symmetry. 
Not only that the tree-level
Dirac masses of neutrinos are forbidden for this additional $\Z_2$ symmetry.
SM gauge group and $\Z_2$ charges of the particles are shown in Tab.~\ref{tab:1}.
\begin{table}[thbp]
\centering {\fontsize{10}{12}
\begin{tabular}{|c||c|c|c|}
\hline Particle & $N_k\, (k = 1,2,3)$ & $ \eta^0 $  & $\eta^+$ 
  \\\hhline{|=#=|=|=|}
$\left(SU(2)_L,U(1)_Y\right)$ & $(\bm{1},0)$ & $(\bm{2},1/2)$ & $(\bm{2},1/2)$
\\\hline
$\Z_2$ & odd~(-) & odd~(-) & odd~(-) \\\hline
\end{tabular}%
} \caption{Additional fields under SM gauge group and $\Z_2$ symmetry}
\label{tab:1}
\end{table}
 
The Lagrangian for the right-handed neutrinos, $N_k$ ($k = 1,2,3$) invariant under both
SM gauge symmetry and $\Z_2$ symmetry can be written as,
\begin{equation}
\mathcal{L}_N=\overline{N_i}i\partial\!\!\!/\!\:P_RN_i
+\left(D_\mu\eta\right)^\dag\left(D^\mu\eta\right)
-\frac{M_i}{2}\overline{N_i\:\!^c}P_RN_i+h_{\alpha
 i}\overline{\ell_\alpha}\eta^\dag P_RN_i+\mathrm{h.c.},
\label{eq:lg}
\end{equation}
where $h_{\alpha k}$, $l_{\alpha}$ and $M_{k}$ represent Yukawa couplings, lepton doublet
and the mass of the right-handed neutrino of type $k$ ($N_{k}$) respectively. 
In our following work $h_{\alpha}$ and $M_{k}$
are chosen to be real without any loss of generality. The invariant scalar
potential containing the Higgs doublet $\Phi$ and the additional SU(2)$_L$ doublet $\eta$
is given by,
\begin{eqnarray}
\mathcal{V}(\phi,\eta)\!\!\!&=&
m_\phi^2\phi^\dag\phi+m_{\eta}^2\eta^\dag\eta
+\frac{\lambda_1}{2}\left(\phi^\dag\phi\right)^2
+\frac{\lambda_2}{2}\left(\eta^\dag\eta\right)^2\nonumber\\
&&\!\!\!\!\!+\lambda_3\left(\phi^\dag\phi\right)\left(\eta^\dag\eta\right)
+\lambda_4\left(\phi^\dag\eta\right)\left(\eta^\dag\phi\right)
+\frac{\lambda_5}{2}\left(\phi^\dag\eta\right)^2+\mathrm{h.c.},
\end{eqnarray}
The tree-level Dirac mass terms for neutrinos can not be generated since 
the vacuum expectation value of the doublet $\eta$ ($\left<\eta\right>$) to 
chosen be zero. 
After electroweak symmetry breaking SM Higgs doublet obtains vacuum expectation value, 
$v = 246$ GeV and the Majorana masses of neutrinos
are generated radiatively via one-loop diagrams with $\eta^0$ and $N_k$ in internal lines.
The model could explain both possibility of scalar ($\eta^0$) and fermion ($N_k$) as DM.
But we choose the mass of one of the three right-handed neutrinos ($N_1$) is considered to be 
lightest among the 
particles added to SM and hence it is a stable candidate of DM. From the forth term of
the Lagrangian in Eq.~\ref{eq:lg} it is clear that the right-handed neutrino interacts only
with the SM lepton doublet and hence leptophilic.

The radiatively generated effective Majorana neutrino masses can be expressed as~\cite{ma},
\begin{equation}
\left(m_\nu\right)_{\alpha\beta}\simeq
\sum_{i=1}^3\frac{2\lambda_5h_{\alpha i}h_{\beta
i}v^2}{(4\pi)^2M_i}I\left(\frac{M_i^2}{M_\eta^2}\right),
\label{eq:neut-mass}
\end{equation}
where $M_\eta^2\simeq
m_\eta^2+\left(\lambda_3+\lambda_4\right)v^2$, $M_i$ are the masses of $\eta$ 
and $N_i$ respectively~\footnote{Masses of the real and imaginary parts of 
$\eta^0$ and $\eta^\pm$ are taken to
 be degenerated for simplicity}. The smallness of the mass term is
guaranteed by the coupling $\lambda_5$. The factor $I\left(x\right)$ can be written as,
\begin{equation}
I\left(x\right)=\frac{x}{1-x}\left(1+\frac{x\log{x}}{1-x}\right).
\end{equation}

Assuming the mass matrix of Eq.~\ref{eq:neut-mass} to be diagonalised using the PMNS matrix
which provides very well explanation for the neutrino oscillation
data, one can find some conditions imposed on $h_{\alpha i}$ as~\cite{Suematsu:2009ww},
\begin{eqnarray}
&&\sum_{k=1}^3\left(2h_{ek}^2\sin 2\theta +2\sqrt 2h_{ek}(h_{\mu k}-h_{\tau k})
\cos 2\theta-(h_{\tau k}-h_{\mu k})^2\sin 2\theta\right)=0, \nonumber \\
&&\sum_{k=1}^3h_{ek}\left(h_{\mu k}+h_{\tau k}\right)=0, \qquad 
\sum_{k=1}^3\left(h_{\mu k}-h_{\tau k}\right)\left(h_{\mu k}
+h_{\tau k}\right)=0.
\label{conditions_hai}
\end{eqnarray}

One of the simple solutions for these conditions on $h_{\alpha i}$ (Eq.~\ref{conditions_hai}) is
achieved by choosing the flavour structure of $h_{\alpha i}$ as,
\begin{equation}
h_{ei}=0, \quad h_{\mu i}=h_{\tau i}; \quad h_{ej}\not=0,
\quad h_{\mu j}=-h_{\tau j}, \,\,\,\, (i\not= j)
\label{yukawa}
\end{equation}
Thus either $i$ or $j$ takes any two values of $k$ (1,2,3). In matrix notation
the structure of the chosen Yukawa couplings of Eq.~\ref{yukawa} can be written as,  
\begin{equation}
h_{\alpha i}=\left(
\begin{array}{ccc}
0   & 0   & h'_3\\
h_1 & h_2 & h_3\\
h_1 & h_2 & -h_3
\end{array}
\right).
\label{eq:fs}
\end{equation}

The Yukawa couplings of Eq.~\ref{eq:fs} imply the values of $\theta_{12}$, $\theta_{23}$
and $\theta_{13}$ to be $\tan^{-1}(\frac{h'_3}{\sqrt{2}h_3})$, $\pi/4$ and 0 respectively.
But from recent observations suggest different values of these mixing angles. 
Then the structure of the matrix will be slightly modified. 
The result of this work will not be vastly modified due to such changes. 

%%%%%%%%%%%%%%%%%%%%%%%%%%%%%%%%%%%%%%%%%%%%%%%%%%%%%%%%%%%%%%%%%%%%%%%%

\section{X-ray line in this framework}

One of the terms in the Lagrangian of this framework that represents the interaction among
the lightest right-handed neutrino ($N_1$), second lightest right-handed
neutrino ($N_2$) and photon is given by~\cite{Schmidt:2012yg},
\begin{equation}
\mathcal{L} = 
i\left(\frac{\mu_{12}}{2}\right)\overline{N_2}\sigma^{\mu\nu}N_1F_{\mu\nu} \,\,\, , 
\label{eq:tmm}
\end{equation}
where $\mu_{12}$ is the coefficient of this interaction and called {\it transition
magnetic moment} between the right-handed neutrinos, $N_1$ and $N_2$. In the 
above $F_{\mu\nu}$ is the so-called electromagnetic field tensor. 
The three-point vertex interaction term of this type is also responsible
in contributing to the inelastic scattering of the right-handed neutrinos
with nucleons via 1-loop processes. 

%%%%%%%%%%%%%%%%%%%
\begin{figure}[cbt]
\begin{center}
\includegraphics[scale=0.65]{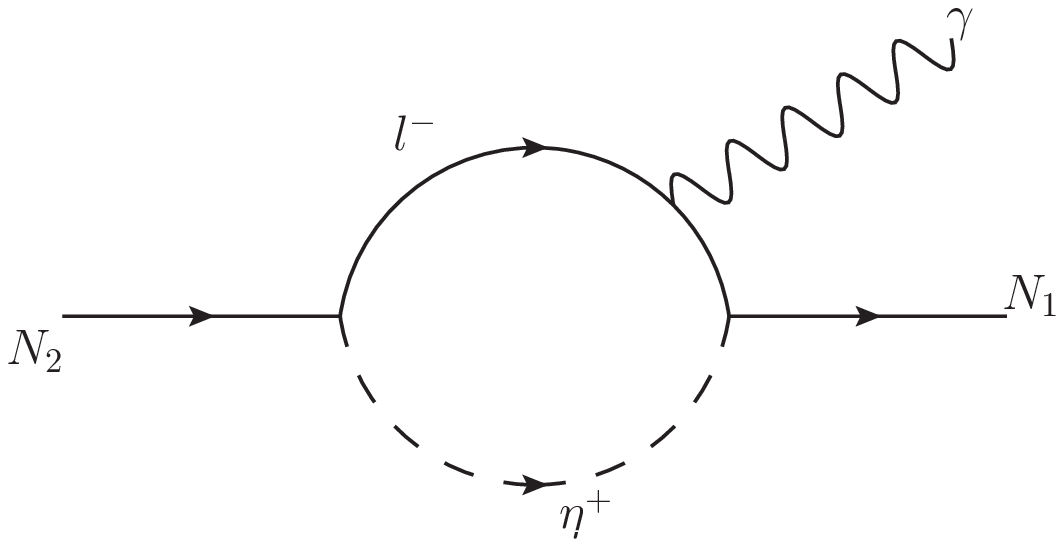}
%%%
\includegraphics[scale=0.65]{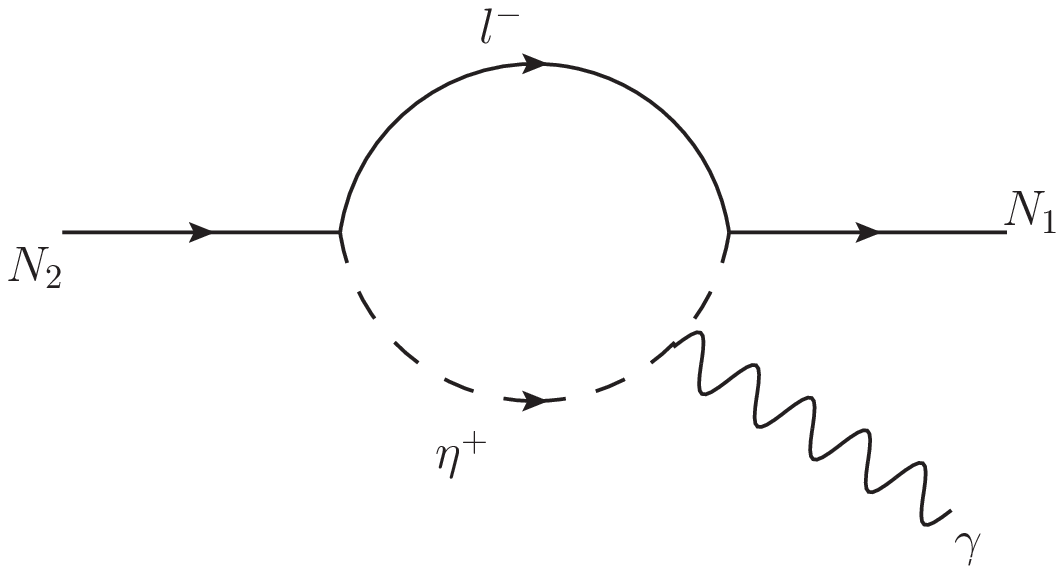}
   \caption{{\it Feynman diagrams showing the decay of second lightest right-handed neutrino, $N_2$
   to lightest right-handed neutrino, $N_1$ and photon ($\gamma$) via radiative processes.}}
   \label{xray}
\end{center}
\end{figure}
%%%%%%%%%%%%%%%%%%%

The X-ray line appears when there is a transition from the state, $N_2$
to $N_1$. The presence of transition magnetic moment solely triggers
such a decay process to occur. The expression of decay width 
for this process can be written as,
\begin{equation}
\Gamma(N_2\to N_1\gamma)=
\frac{\mu_{12}^2}{\pi}\delta^3 \,\,\, ,
\label{ndecay}
\end{equation}
where $\delta = E_{\gamma}$ is the energy of the emitted photon which is nothing
but the mass difference between the lightest and the second lightest right
handed neutrinos present in this framework. The Feynman diagrams responsible
for such process are shown in Fig.~\ref{xray}.

The calculated value of the decay width for the decay process of $N_2$ to $N_1$
and a photon from the observed X-ray line data is $\sim 1.15\times 10^{-52}$ GeV~\cite{Krall:2014dba}.
Thus one can find
from Eq.~\ref{ndecay} that to comply the observed data for X-ray line with the framework of
this model, the absolute value of $\mu_{12}$ should be $\sim 2.9\times 10^{-18}$ GeV$^{-1}$.

The order of the value of $|\mu_{12}|$ is particularly important for studying
the prospects of the direct detection of dark matter. The predicted value of $|\mu_{12}|$
from the recently reported X-ray line data is several orders of magnitude below from the
current reach of various DM direct detection experiments~\cite{Schmidt:2012yg}. 
As the mass of the dark matter
in this model is the lightest right-handed neutrino with heavy mass possibly in the range
from few hundreds of GeV to few thousands of GeV, the direct DM searches should probe
these massive right-handed neutrinos in this mass range.

The expression for $\mu_{12}$ in the present scenario can be written in terms
of model parameters~\cite{Schmidt:2012yg} as,
\begin{equation}
\mu_{12} = - \sum_\alpha
\frac{\mathrm{Im}\left(h_{\alpha 2}^*h_{\alpha 1}\right)e}{2(4\pi)^2M_\eta^2}
2M_1I_\mathrm{m}\left(\frac{M_1^2}{M_\eta^2},\frac{m_{\alpha}^2}{M_\eta^2}\right),
\label{eq:mu12}\\
\end{equation}
$m_{\alpha}$ is the mass eigenvalue of ordinary neutrino of flavour $\alpha$, $e$
is the electric charge of proton. The term $\mathrm{Im}\left(h_{\alpha 2}^*h_{\alpha 1}\right)$ 
in Eqn.~\ref{eq:mu12} is related to the phase difference, $\xi$ between the Yukawa couplings
$h_{\alpha 2}$ and $h_{\alpha 1}$ for flavour $\alpha$.
% For the matrix of Yukawa couplings of Eq.~\ref{eq:fs} 
% $$\sum_\alpha 
% \mathrm{Im}\left(h_{\alpha 2}^*h_{\alpha 1}\right) = 
% 2\times\mathrm{Im}\left(h_{\alpha 2}^*h_{\alpha 1}\right).$$ 
For the matrix of Yukawa couplings of Eq.~\ref{eq:fs} the value of the factor,
$\mathrm{Im}\left(h_{\alpha 2}^*h_{\alpha 1}\right)$ is zero
for one flavour and contributes equally for the remaining flavours.  
In the above the function $I_{\mathrm{m}}$ comes from loop
integral and can be expressed as, 
\begin{equation}
I_{\mathrm{m}}(x,y) = - \int_0^1\frac{z(1-z)}{xz^2-(1+x-y)z+1} dz.
\end{equation}
%

%%%%%%%%%%%%%%%%%%%
\begin{figure}[tbc]
\begin{center}
\includegraphics[scale=0.55,angle=-90]{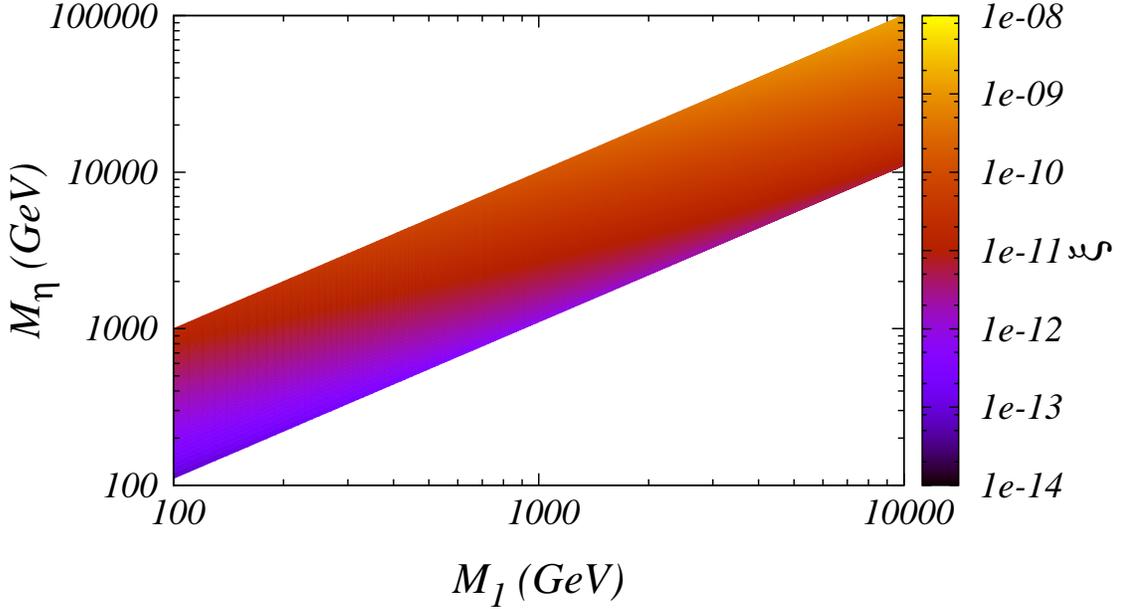}
   \caption{ {\it The allowed parameter space consisting of $M_1$, $M_{\eta}$ and $\xi$ 
   consistent with the recently reported 3.5 keV X-ray line data.
   The value of ratio of the mass of $N_1$ to that of $\eta$ is chosen to be 
   within 10.0, i.e., $1.0 < M_{\eta}/M_{1} \leq 10.0$ in this plot.
   The considered range of $M_1$ is from $10^2$ GeV to $10^4$ GeV.
   The phase factor $\xi$ are shown by the colour index
   where $\xi$ varies from blue coloured region to yellow 
   region as its value increases. See text for more details.}
   }
   \label{param}
\end{center}
\end{figure}
%%%%%%%%%%%%%%%%%%%

Considering masses of ordinary neutrinos are negligible with respect to that
of $\eta$, i.e., $m_{\alpha} \ll M_{\eta}$, the allowed parameter space for
the model parameters, $M_1$, $M_{\eta}$ and $\xi$ is obtained from the
computed value of $|\mu_{12}|$ from 3.5 keV X-ray line data.~\footnote{The 
mass of ordinary neutrino is several orders of magnitude smaller than
the mass of doublet scalar $\eta$ which is few hundreds of GeV or more
in this framework and hence the ratio, $\frac{m_{\alpha}}{M_{\eta}}$ is $\ll 1$} 
The plot showing the variation of the parameters 
constrained from observed X-ray line data is shown in Fig.~\ref{param}. 
In this plot the ratio ($r$)
of $M_{\eta}$ to $M_{1}$ is taken to be between 1.0 to 10.0, i.e., $1.0 < M_{\eta}/M_{1} \leq 10.0$.
The range of the constrained values of the phase factor, $\xi$
for those mass ratios ($1.0 < r\le 10.0$) spanning from $\sim 10^{-14}$ to $\sim 10^{-8}$.
The situation would have been slightly modified if one incorporate the 
precise values of mixing angles (for example, non-zero $\theta_{13}$). 
The Yukawa matrix structure is then modified and the phase factor
for each flavour $\alpha$ will be different in general. But it can be shown  
that for such cases the the order of the sum of the phase factors 
will be almost of similar order that has been obtained in this case.
The phase factor determines the coannihilation of $N_1-N_2$ and the effective interaction of
right-handed neutrino DM with nuclei.
The result shows the values of phase factor $\xi$ with 
much smaller orders for the considered mass range than expected to be give 
signatures of direct detection.
Hence the coannihilation channels and the DM-nuclei
interaction is much lowered from the computed value of $\xi$ constrained
by the 3.5 keV X-ray line data.
Thus the possibility of direct detection of dark matter
in this framework is suppressed by few orders
from the reach of ongoing direct DM search experiments.

\section{Summary and Conclusion}

We have shown that the radiative neutrino mass model can explain the observed 3.5 keV X-ray line signal
from the data of various galaxy clusters and Andromeda galaxy (M31). 
This model can accommodate naturally both neutrino mass and stable cold dark matter candidate.
The small mass difference between
the lightest and the second lightest right-handed neutrino have been considered to produce the energy
of the X-ray signal. Thus the transition from $N_2 \to N_1 + \gamma$ due to transition magnetic moment
via radiative processes involving leptons and charged scalar in internal lines can
naturally accommodate all the requirements for the X-ray line signal. The value of the
transition magnetic moment ($\mu_{12}$) for such an observed signal is estimated to be few orders
of magnitude smaller than the reach of recent DM direct direct detection experimental limits 
sustaining the possibility of the cold DM candidate in this model to be detected directly. 
The other parameters of this model, namely masses of lightest right-handed neutrino ($N_1$),
doublet scalar ($\eta$) and phase factor ($\xi$) between Yukawa couplings, $h_1$ and $h_2$ 
are further constrained from the observed X-ray line data. 
A very small but non-zero value of the phase difference between Yukawa couplings, $h_1$ and $h_2$
have been predicted.
Also the coannihilation between $N_1$ and $N_2$ is reduced and the s-wave contribution of dark
matter annihilation cross section is calculated to be reduced.
Finally the analysis performed here for this model framework would be viable for any 
DM signal in this energy regime. 
In addition the dark matter candidate (lightest right-handed neutrino), 
being leptophilic and massive, can potentially explain AMS-02 positron excess.

% 
% \fi
% 
% \newpage
% %%%%%%%%%%%%%%%%%%%%%%%%%%%%%%%%%%
% \vspace{0.5cm}
% \hspace{0.2cm} {\bf Acknowledgments}
\section*{Acknowledgments}
\vspace{0.5cm}
I would like to thank Debasish Majumdar for helpful suggestions.
I also want to acknowledge Department of Atomic Energy (DAE, Govt. of India) for financial
assistance.
%%%%%%%%%%%%%%%%%%%%%%%%%%%%%%%%%%


\begin{thebibliography}{99}

%%%%%%%%%% DM expt. %%%%%%%%%%%%%%
\bibitem{Ade:2013lta} 
  P.~A.~R.~Ade {\it et al.}  [ Planck Collaboration],
  %``Planck 2013 results. XVI. Cosmological parameters,''
  arXiv:1303.5076 [astro-ph.CO].
  %%CITATION = ARXIV:1303.5076;%%
  
\bibitem{Begeman:1991iy}
  K.~G.~Begeman, A.~H.~Broeils and R.~H.~Sanders,
  %``Extended rotation curves of spiral galaxies: Dark haloes and
	%modified
  %dynamics,''
  Mon.\ Not.\ Roy.\ Astron.\ Soc.\  {\bf 249}, 523 (1991).
  %%CITATION = MNRAA,249,523;%%



%dark matter
%\cite{Komatsu:2010fb}
\bibitem{Komatsu:2010fb}
  E.~Komatsu {\it et al.}  [WMAP Collaboration],
  %``Seven-Year Wilkinson Microwave Anisotropy Probe (WMAP)
	%Observations:
  %Cosmological Interpretation,''
  Astrophys.\ J.\ Suppl.\  {\bf 192}, 18 (2011)
  [arXiv:1001.4538 [astro-ph.CO]].
  %%CITATION = APJSA,192,18;%%


%\cite{Massey:2007wb}
\bibitem{Massey:2007wb}
  R.~Massey {\it et al.},
  %``Dark matter maps reveal cosmic scaffolding,''
  Nature {\bf 445}, 286 (2007)
  [arXiv:astro-ph/0701594].
  %%CITATION = NATUA,445,286;%%

 \bibitem{scalar_singlet} V. Silveira and A. Zee, Phys. Lett. B {\bf 161}, 136 (1985).

 \bibitem{veltman} M.~J.~G.~Veltman, F.~J.~Yndurain, Nucl.\ Phys.\ B{\bf 325}, 1 (1989).

\bibitem{Burgess:2000yq}
  C.~P.~Burgess, M.~Pospelov, T.~ter Veldhuis,
  %``The Minimal model of nonbaryonic dark matter: A Singlet scalar,''
  Nucl.\ Phys.\  {\bf B619}, 709-728 (2001).
  [hep-ph/0011335].
  
  \bibitem{kaluza} H.~-C.~Cheng, J.~L.~Feng and K.~T.~Matchev,
  %``Kaluza-Klein dark matter,''
  Phys.\ Rev.\ Lett.\  {\bf 89}, 211301 (2002)
  [hep-ph/0207125]; G.~Servant and T.~M.~P.~Tait,
  %``Is the lightest Kaluza-Klein particle a viable dark matter candidate?,''
  Nucl.\ Phys.\ B {\bf 650}, 391 (2003)
  [hep-ph/0206071].

 \bibitem{triplet} T.~Araki, C.~Q.~Geng and K.~I.~Nagao,
%   %``Dark Matter in Inert Triplet Models,''
   Phys.\ Rev.\ D {\bf 83}, 075014 (2011)
   [arXiv:1102.4906 [hep-ph]].

\bibitem{Modak:2012wk}
  K.~P.~Modak and D.~Majumdar,
  %``Gamma Ray and Neutrino Flux from Annihilation of Neutralino Dark Matter at Galactic Halo Region in mAMSB Model,''
  J.\  Phys.\  G: Nucl.\  Part.\  Phys.\  {\bf 40}, 075201 (2013)
  [arXiv:1205.1996 [hep-ph]].
  %%CITATION = ARXIV:1205.1996;%%

\bibitem{smssm} R.~Kappl, M.~Ratz and M.~W.~Winkler,
   %``Light dark matter in the singlet-extended MSSM,''
   Phys.\ Lett.\ B {\bf 695}, 169 (2011)
   [arXiv:1010.0553 [hep-ph]].
\bibitem{axion} L.~D.~Duffy and K.~van Bibber,
  %``Axions as Dark Matter Particles,''
  New J.\ Phys.\  {\bf 11}, 105008 (2009)
  [arXiv:0904.3346 [hep-ph]].

\bibitem{sing_ferm} Y.~G.~Kim, K.~Y.~Lee and S.~Shin,
  %``Singlet fermionic dark matter,''
  JHEP {\bf 0805}, 100 (2008)
  [arXiv:0803.2932 [hep-ph]].


\bibitem{idm} L.~Lopez Honorez, E.~Nezri, J.~F.~Oliver and M.~H.~G.~Tytgat,
  %``The Inert Doublet Model: An Archetype for Dark Matter,''
  JCAP {\bf 0702}, 028 (2007)
  [hep-ph/0612275].


\bibitem{Cirelli}
  M.~Cirelli, G.~Corcella, A.~Hektor, G.~Hutsi, M.~Kadastik, P.~Panci, M.~Raidal and F.~Sala {\it et al.},
  %``PPPC 4 DM ID: A Poor Particle Physicist Cookbook for Dark Matter Indirect Detection,''
  JCAP {\bf 1103}, 051 (2011)
  [Erratum-ibid.\  {\bf 1210}, E01 (2012)]
  [arXiv:1012.4515 [hep-ph]].
 
 \bibitem{Hooper:2012sr} 
  D.~Hooper, C.~Kelso and F.~S.~Queiroz,
  %``Stringent and Robust Constraints on the Dark Matter Annihilation Cross Section From the Region of the Galactic Center,''
  Astropart.\ Phys.\  {\bf 46}, 55 (2013)
  [arXiv:1209.3015 [astro-ph.HE]].
  %%CITATION = ARXIV:1209.3015;%%
 \bibitem{Modak:2013jya} 
  K.~P.~Modak, D.~Majumdar and S.~Rakshit,
  %``A Possible Explanation of Low Energy $\gamma$-ray Excess from Galactic Centre and Fermi Bubble by a Dark Matter Model with Two Real Scalars,''
  arXiv:1312.7488 [hep-ph].
 

 %\cite{Bulbul:2014sua}
\bibitem{Bulbul:2014sua} 
  E.~Bulbul, M.~Markevitch, A.~Foster, R.~K.~Smith, M.~Loewenstein and S.~W.~Randall,
  %``Detection of An Unidentified Emission Line in the Stacked X-ray spectrum of Galaxy Clusters,''
  arXiv:1402.2301 [astro-ph.CO].
  %%CITATION = ARXIV:1402.2301;%%
  %6 citations counted in INSPIRE as of 05 Mar 2014 
  
  %\cite{Boyarsky:2014jta}
\bibitem{Boyarsky:2014jta} 
  A.~Boyarsky, O.~Ruchayskiy, D.~Iakubovskyi and J.~Franse,
  %``An unidentified line in X-ray spectra of the Andromeda galaxy and Perseus galaxy cluster,''
  arXiv:1402.4119 [astro-ph.CO].
  %%CITATION = ARXIV:1402.4119;%%
  %5 citations counted in INSPIRE as of 05 Mar 2014
  
  
  %\cite{Ishida:2014dlp}
\bibitem{Ishida:2014dlp} 
  H.~Ishida, K.~S.~Jeong and F.~Takahashi,
  %``7 keV sterile neutrino dark matter from split flavor mechanism,''
  arXiv:1402.5837 [hep-ph].
  %%CITATION = ARXIV:1402.5837;%%
  %3 citations counted in INSPIRE as of 05 Mar 2014
  
  %\cite{Finkbeiner:2014sja}
\bibitem{Finkbeiner:2014sja} 
  D.~P.~Finkbeiner and N.~Weiner,
  %``An X-Ray Line from eXciting Dark Matter,''
  arXiv:1402.6671 [hep-ph].
  %%CITATION = ARXIV:1402.6671;%%
  %1 citations counted in INSPIRE as of 05 Mar 2014
  
  %\cite{Higaki:2014zua}
\bibitem{Higaki:2014zua} 
  T.~Higaki, K.~S.~Jeong and F.~Takahashi,
  %``The 7 keV axion dark matter and the X-ray line signal,''
  arXiv:1402.6965 [hep-ph].
  %%CITATION = ARXIV:1402.6965;%%
  %2 citations counted in INSPIRE as of 05 Mar 2014

\bibitem{Jaeckel:2014qea} 
  J.~Jaeckel, J.~Redondo and A.~Ringwald,
  %``A 3.55 keV hint for decaying axion-like particle dark matter,''
  arXiv:1402.7335 [hep-ph].
  %%CITATION = ARXIV:1402.7335;%%  
  
  %\cite{Lee:2014xua}
\bibitem{Lee:2014xua} 
  H.~M.~Lee, S.~C.~Park and W.~-I.~Park,
  %``Cluster X-ray line at $3.5\,{\rm keV}$ from axion-like dark matter,''
  arXiv:1403.0865 [astro-ph.CO].
  %%CITATION = ARXIV:1403.0865;%%
  
%\cite{Kong:2014gea}
\bibitem{Kong:2014gea} 
  K.~Kong, J.~-C.~Park and S.~C.~Park,
  %``X-ray line signal from 7 keV axino dark matter decay,''
  arXiv:1403.1536 [hep-ph].
  %%CITATION = ARXIV:1403.1536;%%

\bibitem{Choi:2014tva} 
  K.~-Y.~Choi and O.~Seto,
  %``X-ray line signal from decaying axino warm dark matter,''
  arXiv:1403.1782 [hep-ph].

\bibitem{Baek:2014qwa} 
  S.~Baek and H.~Okada,
  %``7 keV Dark Matter as X-ray Line Signal in Radiative Neutrino Model,''
  arXiv:1403.1710 [hep-ph].

\bibitem{Tsuyuki:2014aia} 
  T.~Tsuyuki,
  %``Neutrino masses, leptogenesis, and sterile neutrino dark matter,''
  arXiv:1403.5053 [hep-ph].

\bibitem{Bezrukov:2014nza} 
  F.~Bezrukov and D.~Gorbunov,
  %``Relic Gravity Waves and 7 keV Dark Matter from a GeV scale inflaton,''
  arXiv:1403.4638 [hep-ph].

\bibitem{Kolda:2014ppa} 
  C.~Kolda and J.~Unwin,
  %``X-ray lines from R-parity violating decays of keV sparticles,''
  arXiv:1403.5580 [hep-ph].

\bibitem{Allahverdi:2014dqa} 
  R.~Allahverdi, B.~Dutta and Y.~Gao,
  %``keV Photon Emission from Light Nonthermal Dark Matter,''
  arXiv:1403.5717 [hep-ph].

\bibitem{Queiroz:2014yna} 
  F.~S.~Queiroz and K.~Shinha,
  %``The Poker Face of the Majoron Dark Matter Model: LUX to keV Line,''
  arXiv:1404.1400 [hep-ph].

\bibitem{Babu:2014pxa} 
  K.~S.~Babu and R.~N.~Mohapatra,
  %``7 keV Scalar Dark Matter and the Anomalous Galactic X-ray Spectrum,''
  arXiv:1404.2220 [hep-ph].

\bibitem{Dudas:2014ixa} 
  E.~Dudas, L.~Heurtier and Y.~Mambrini,
  %``Generating X-ray lines from annihilating dark matter,''
  arXiv:1404.1927 [hep-ph].

\bibitem{Demidov:2014hka} 
  S.~V.~Demidov and D.~S.~Gorbunov,
  %``SUSY in the sky or keV signature of sub-GeV gravitino dark matter,''
  arXiv:1404.1339 [hep-ph].

\bibitem{Ko:2014xda} 
  P.~Ko, Z.~kang, T.~Li and Y.~Liu,
  %``Natural $X$-ray Lines from the Low Scale Supersymmetry Breaking,''
  arXiv:1403.7742 [hep-ph].

\bibitem{Bomark:2014yja} 
  N.~-E.~Bomark and L.~Roszkowski,
  %``The 3.5 keV X-ray line from decaying gravitino dark matter,''
  arXiv:1403.6503 [hep-ph].

\bibitem{Liew:2014gia} 
  S.~P.~Liew,
  %``Axino dark matter in light of an anomalous X-ray line,''
  arXiv:1403.6621 [hep-ph].

\bibitem{Krall:2014dba} 
  R.~Krall, M.~Reece and T.~Roxlo,
  %``Effective field theory and keV lines from dark matter,''
  arXiv:1403.1240 [hep-ph].

\bibitem{Aisati:2014nda} 
  C.~ïm.~E.~Aisati, T.~Hambye and T.~Scarna,
  %``Can a millicharged dark matter particle emit an observable gamma-ray line?,''
  arXiv:1403.1280 [hep-ph].

\bibitem{Frandsen:2014lfa} 
  M.~Frandsen, F.~Sannino, I.~M.~Shoemaker and O.~Svendsen,
  %``X-ray Lines from Dark Matter: The Good, The Bad, and The Unlikely,''
  arXiv:1403.1570 [hep-ph].
  
\bibitem{Fukuda:1998mi}
  Y.~Fukuda {\it et al.}  [Super-Kamiokande Collaboration],
  %``Evidence for oscillation of atmospheric neutrinos,''
  Phys.\ Rev.\ Lett.\  {\bf 81}, 1562 (1998)
  [arXiv:hep-ex/9807003].
  %%CITATION = PRLTA,81,1562;%%

%\cite{Ahmad:2002jz}
\bibitem{Ahmad:2002jz}
  Q.~R.~Ahmad {\it et al.}  [SNO Collaboration],
  %``Direct evidence for neutrino flavor transformation from neutral
	%current
  %interactions in the Sudbury Neutrino Observatory,''
  Phys.\ Rev.\ Lett.\  {\bf 89}, 011301 (2002)
  [arXiv:nucl-ex/0204008].
  %%CITATION = PRLTA,89,011301;%%


%\cite{Araki:2004mb}
\bibitem{Araki:2004mb}
  T.~Araki {\it et al.}  [KamLAND Collaboration],
  %``Measurement of neutrino oscillation with KamLAND: Evidence of
	%spectral
  %distortion,''
  Phys.\ Rev.\ Lett.\  {\bf 94}, 081801 (2005)
  [arXiv:hep-ex/0406035].
  %%CITATION = PRLTA,94,081801;%%

%\cite{Adamson:2008zt}
\bibitem{Adamson:2008zt}
  P.~Adamson {\it et al.}  [MINOS Collaboration],
  %``Measurement of Neutrino Oscillations with the MINOS Detectors in
	%the NuMI
  %Beam,''
  Phys.\ Rev.\ Lett.\  {\bf 101}, 131802 (2008)
  [arXiv:0806.2237 [hep-ex]].
  %%CITATION = PRLTA,101,131802;%%
  


\bibitem{ma}
E. Ma, 
Phys.\ Rev.\ D {\bf 73}, 077301 (2006) 
[arXiv:hep-ph/0601225]


\bibitem{Kubo:2006yx}
  J.~Kubo, E.~Ma and D.~Suematsu,
  %``Cold Dark Matter, Radiative Neutrino Mass, mu ---> e gamma, and
  %Neutrinoless Double Beta Decay,''
  Phys.\ Lett.\  B {\bf 642}, 18 (2006)
  [arXiv:hep-ph/0604114].
  %%CITATION = PHLTA,B642,18;%%

%\cite{Kajiyama:2006ww}
\bibitem{Kajiyama:2006ww}
  Y.~Kajiyama, J.~Kubo and H.~Okada,
  %``D(6) Family Symmetry and Cold Dark Matter at LHC,''
  Phys.\ Rev.\  D {\bf 75}, 033001 (2007)
  [arXiv:hep-ph/0610072].
  %%CITATION = PHRVA,D75,033001;%%

%\cite{Suematsu:2009ww}
\bibitem{Suematsu:2009ww}
  D.~Suematsu, T.~Toma and T.~Yoshida,
  %``Reconciliation of CDM abundance and mu ---> e gamma in a radiative
	%seesaw
  %model,''
  Phys.\ Rev.\  D {\bf 79}, 093004 (2009)
  [arXiv:0903.0287].
  %%CITATION = PHRVA,D79,093004;%%

%\cite{Suematsu:2010gv}
\bibitem{Suematsu:2010gv}
  D.~Suematsu, T.~Toma and T.~Yoshida,
  %``Enhancement of the annihilation of dark matter in a radiative
	%seesaw
  %model,''
  Phys.\ Rev.\  D {\bf 82}, 013012 (2010)
  [arXiv:1002.3225].
  %%CITATION = PHRVA,D82,013012;%%

%\cite{Sierra:2008wj}
\bibitem{Sierra:2008wj}
  D.~Aristizabal Sierra, J.~Kubo, D.~Restrepo, D.~Suematsu and
	O.~Zapata,
  %``Radiative seesaw: Warm dark matter, collider and lepton flavour
	%violating
  %signals,''
  Phys.\ Rev.\  D {\bf 79}, 013011 (2009)
  [arXiv:0808.3340].
  %%CITATION = PHRVA,D79,013011;%%
\bibitem{Kajiyama:2011fe}
  Y.~Kajiyama, H.~Okada and T.~Toma,
  %``Direct and Indirect Detection of Dark Matter in D6 Flavor Symmetric
  %Model,''
  Eur.\ Phys.\ J.\  C {\bf 71}, 1688 (2011)
  [arXiv:1104.0367].
  %%CITATION = EPHJA,C71,1688;%%

%\cite{Kajiyama:2011fx}
\bibitem{Kajiyama:2011fx}
  Y.~Kajiyama, H.~Okada and T.~Toma,
  %``A light Scalar Dark Matter for CoGeNT and DAMA in D_6 Flavor
	%Symmetric
  %Model,''
  arXiv:1109.2722.
  %%CITATION = ARXIV:1109.2722;%%


\bibitem{Schmidt:2012yg} 
  D.~Schmidt, T.~Schwetz and T.~Toma,
  %``Direct Detection of Leptophilic Dark Matter in a Model with Radiative Neutrino Masses,''
  Phys.\ Rev.\ D {\bf 85}, 073009 (2012)
  [arXiv:1201.0906 [hep-ph]].
  


\end{thebibliography}
\end{document}